\documentclass[journal]{IEEEtran}
\IEEEoverridecommandlockouts
\usepackage{cite}

\usepackage{cite,setspace,bm,enumerate}
\usepackage{amsmath,amssymb,amsfonts}
\usepackage{subeqnarray,subfigure}
\usepackage{algorithmic}
\usepackage{graphicx}
\usepackage{epstopdf}
\usepackage{textcomp}
\usepackage{xcolor}
\usepackage{verbatim}
\usepackage{booktabs}
\usepackage[linesnumbered,ruled,vlined]{algorithm2e}
\usepackage{amsmath}
\def\BibTeX{{\rm B\kern-.05em{\sc i\kern-.025em b}\kern-.08em
		T\kern-.1667em\lower.7ex\hbox{E}\kern-.125emX}}

\begin{document}
	
	\title{Energy-Efficient UAV Swarm Assisted MEC with Dynamic Clustering and Scheduling}

	\author{
		
		Jialiuyuan Li$^{\ast}$, Jiayuan Chen$^{\ast}$, Changyan Yi,$^{\ast}$, Tong Zhang$^{\ast}$, Kun Zhu$^{\ast}$ and Jun Cai$^{\dagger}$\\
		\IEEEauthorblockA{\text{\small $^{\ast}$College of Computer Science and Technology, Nanjing University of Aeronautics and Astronautics, Nanjing, China} \\
\small $^{\dagger}$Department of Electrical and Computer Engineering, Concordia University, Montr\'{e}al, QC, H3G 1M8, Canada \\
\text{\small Email: \{jialiuyuan.li, jiayuan.chen, changyan.yi, zhangt, zhukun\}@nuaa.edu.cn,  jun.cai@concordia.ca}\\
	}
}

	\maketitle

		\begin{abstract}
			
			In this paper, the energy-efficient unmanned aerial vehicle (UAV) swarm assisted mobile edge computing (MEC) with dynamic clustering and scheduling is studied. In the considered system model, UAVs are divided into multiple swarms, with each swarm consisting of a leader UAV and several follower UAVs to provide computing services to end-users. Unlike existing work, we allow UAVs to dynamically cluster into different swarms, i.e., each follower UAV can change its leader based on the time-varying spatial positions, updated application placement, etc. in a dynamic manner. Meanwhile, UAVs are required to dynamically schedule their energy replenishment, application placement, trajectory planning and task delegation. With the aim of maximizing the long-term energy efficiency of the UAV swarm assisted MEC system, a joint optimization problem of dynamic clustering and scheduling is formulated. Taking into account the underlying cooperation and competition among intelligent UAVs, we further reformulate this optimization problem as a combination of a series of strongly coupled multi-agent stochastic games, and then propose a novel reinforcement learning-based UAV swarm dynamic coordination (RLDC) algorithm for obtaining the equilibrium. Simulations are conducted to evaluate the performance of the RLDC algorithm and demonstrate its superiority over counterparts.
			
		\end{abstract}

\vspace{-0.2em}
	
	\section{Introduction}\label{Section_Intro}
	
	\IEEEPARstart{R}{ecently}, unmanned aerial vehicle (UAV) assisted mobile edge computing (MEC) \cite{57,53} has attracted significant attentions due to its high mobility, flexible coverage and rapid deployment in providing fast-responsive supplementary computing services to end-users (e.g., IoT devices). Furthermore, by forming into swarms (each of which consists of a leader and multiple followers\cite{67}), UAV swarm assisted MEC can further improve the collaboration among UAVs for enhancing the service quality, and thus has become a popular trend for future applications\cite{4}.

	Although UAV swarm assisted MEC is envisioned as a lightweight and highly efficient paradigm, it faces several inherent challenges: i) since the MEC service demands of IoT devices vary dynamically, if UAV swarms are predetermined with fixed clustering, the computing workloads among different swarms may be severely unbalanced; ii) UAVs (especially the leaders) are battery-constrained and have to fly to the depot for energy replenishment if necessary, meaning that their swarm formations cannot be maintained long-term static; iii) the limited storage capacities of UAVs (both leaders and followers) impede their abilities to store all applications to fulfill diverse task requirements of IoT devices, indicating that they have to help with each other through task delegations (particularly within the swarm). Recent research efforts in this area include cooperative trajectory planning \cite{60,61} and collaborative task delegation \cite{62,75}, etc. Nevertheless, there are still some critical issues, especially how UAV swarms can cater to dynamic service requirements of IoT devices, and how UAV swarms can be dynamically clustered based on their spatial positions and updated application placement, which are of great importance but have not yet been well investigated.
	
	In this paper, we study a joint optimization problem of dynamic clustering and scheduling for UAV swarm assisted MEC to maximize the long-term energy efficiency of the system. Specifically, in the considered model, the following decisions are made within each time slot: i) each leader UAV determines whether it should return to the depot for refueling energy and updating the installed applications or the next target service region of its leaded swarm; ii) each follower UAV determines which leader UAV to follow (i.e., the associated swarm), the trajectory in its swarm, and whether to delegate certain tasks to the leader. Since all UAVs are intelligent, they are allowed to make individual decisions, potentially leading to the cooperation and competition among them. To this end, we reformulate the joint optimization problem as a series of complex multi-agent stochastic games, i.e., energy replenishment stochastic game (ERSG), application planning stochastic game (APSG), leader UAV trajectory planning stochastic game (LTSG), dynamic clustering stochastic game (DCSG), follower UAV trajectory planning stochastic game (FTSG), and task delegation stochastic game (TDSG). After analyzing their properties, we then propose a novel reinforcement learning-based UAV swarm dynamic coordination (RLDC) algorithm to obtain the corresponding equilibriums.

The main contributions of this paper are as follows.

\begin{itemize}
	\item[$\bullet$]A joint optimization problem of dynamic clustering and scheduling in UAV swarm assisted MEC is formulated, where the objective is to maximize the long-term energy efficiency of all UAVs (including leaders and followers).
	\item[$\bullet$]Observing the cooperation and competition among UAVs, we reformulate the optimization problem as a series of coupled multi-agent stochastic games, and then propose a novel algorithm, called RLDC, to obtain the corresponding equilibriums.
	\item[$\bullet$]Extensive simulations are conducted to show the superiority of the proposed RLDC algorithm over counterparts.
\end{itemize}

The rest of this paper is organized as follows: Section II introduces the system model and problem formulation. In Section III, a problem reformulation based on multi-agent stochastic game is proposed and analyzed, along with the developed RLDC algorithm. Simulation results are provided in Section IV, followed by the conclusion in Section V.
\section{System Model and Problem Formulation}\label{SectionSYS}
\subsection{Network Model}

\begin{figure}[t]  
	\centering
	\includegraphics[width=3.6in]{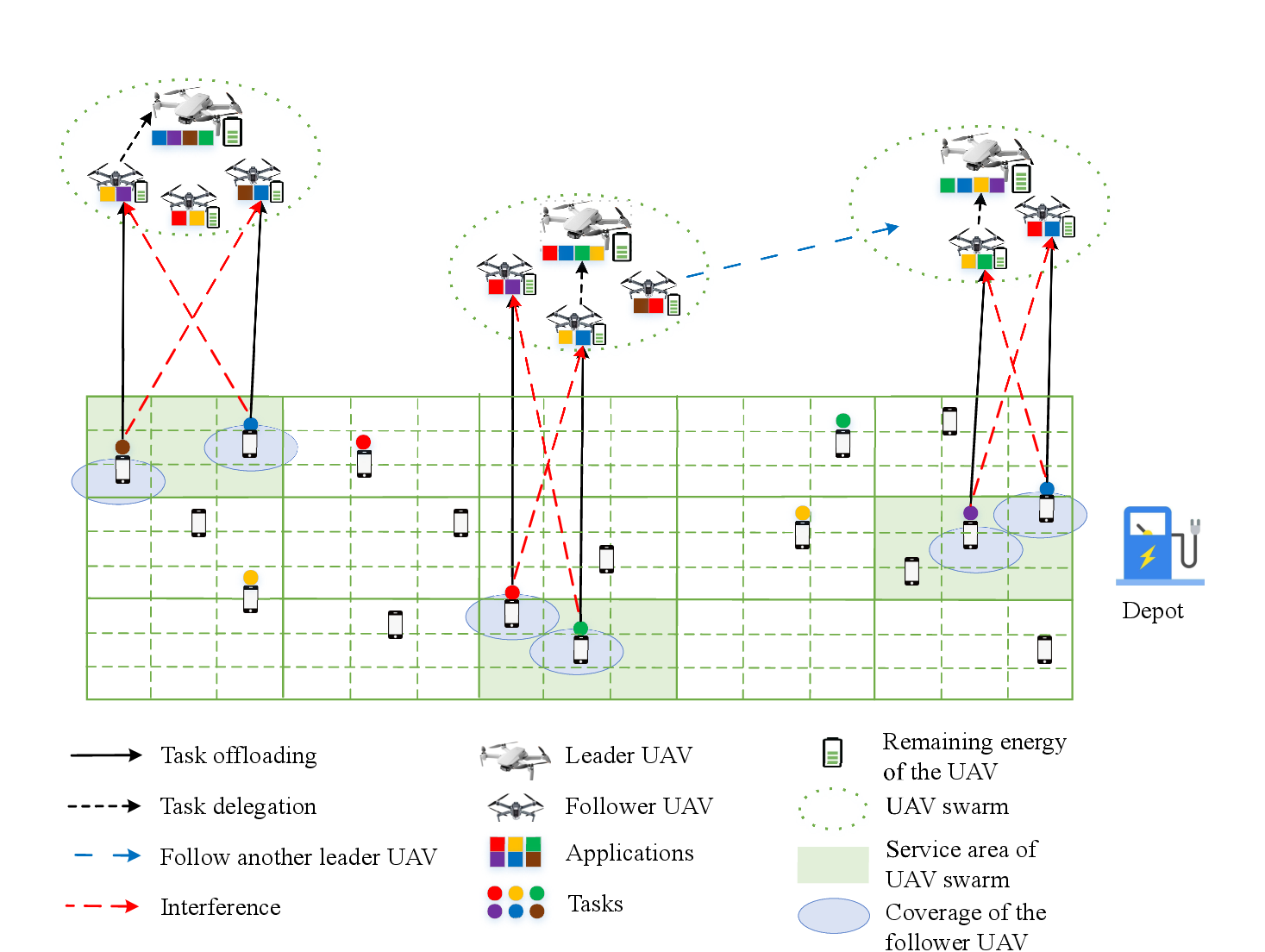}\\
	\caption{An illustration of the considered UAV swarm assisted MEC.}\label{model}
\end{figure}
Consider a UAV swarm assisted MEC system deployed in a target region, as illustrated in Fig. \ref{model}. The system consists of a group of leader UAVs, denoted as $\mathcal{M}$ with $|\mathcal{M}| = M$, a group of follower UAVs, denoted as $\mathcal{N}$ with $|\mathcal{N}| = N$, and a set of IoT devices scattered randomly on the ground, denoted as $\mathcal{K}$ with $|\mathcal{K}| = K$. At the edge of the target region, a depot is deployed to serve leader UAVs with energy replenishment and application placement update through wired connections. A time-slotted operation framework is studied, which is characterized by $t\in\{1,2,...,T\}$. The target region is uniformly divided into large grids with side length $l$ to limit the activities scope of each swarm. Meanwhile, the large grids are further uniformly divided into small grids with side length $q$ to specify the activities of follower UAVs. Similar to \cite{43}, the downlink transmission range of each follower UAV is assumed to be $\sqrt{2}q/2$, such that it can cover a small grid for computation outcome feedback. At any time slot, each small grid is limited to be covered by only one follower UAV to avoid collisions, where each follower UAV provides supplementary computing services to multiple IoT devices simultaneously within its downlink transmission range. We denote the set of IoT devices served by follower UAV $n \in \mathcal{N}$ as $\mathcal{G}_n$, and each IoT devices $k \in \mathcal{K}$ uses the same frequency band $B$ for uplink communications. Then, the signal-to-interference-plus-noise ratio (SINR) at follower UAV $n$ with regard to the uplink communication of IoT device $k$, and the SINR at leader UAV $m$ with regard to the uplink communication of follower UAV $n$ at time slot $t$, can be expressed as:
\begin{equation}
	\begin{array}{l}
		\resizebox{0.65\hsize}{!}{$\gamma_{n,k}(t)=\frac{p^I_k10^{\frac{-\lambda_{n,k}(t)}{10}}}{\sum\limits_{i\in\mathcal{G}_n\backslash k} p^I_i10^{\frac{-\lambda_{n,k}(t)}{10}}+\varpi}$},
	\end{array}
\end{equation}
and
\begin{equation}
	\begin{array}{l}
		\resizebox{0.5\hsize}{!}{$\gamma_{m,n}(t)=\frac{p^F_n10^{\frac{-\lambda_{m,n}(t)}{10}}}{\varpi}$},
	\end{array}
\end{equation}
respectively. $p^I_k$ and $p^F_n$ denote the transmission power of IoT device $k$ and follower UAV $n$, respectively. $\varpi$ indicates the power spectral density of noise. Following the literature \cite{61}, $\lambda_{m,n}(t)$ and $\lambda_{n,k}(t)$ indicate the path loss between leader UAV $m$ and follower UAV $n$, and the path loss between follower UAV $n$ and IoT device $k$ at time slot $t$, respectively. Let $\mathcal{C}=\{1,2,...,C\}$ denote the set of task type. Based on above expressions, the time of IoT device $k\in\mathcal{G}_n$ offloading task $c\in\mathcal{C}$ to follower UAV $n$, and the time of follower UAV $n$ delegating task $c$ to leader UAV $m$ at time slot $t$ can be written as:
\begin{equation}
	T^{off}_{n,k,c}(t)=\frac{\upsilon_{k,c}(t) \kappa_{k,c}}{B\log_2(1+\gamma_{n,k}(t))},
\end{equation}
and
\begin{equation}
	T^{dele}_{m,n,c}(t)=\frac{(1-\varepsilon_m(t))\delta_{m,n}(t)\phi_{m,n,c}(t) \kappa_{k,c}}{B\log_2(1+\gamma_{m,n}(t))},
\end{equation}
respectively. $\kappa_{k,c}$ indicates the size of task $c$ offloaded from IoT device $k$. $\upsilon_{k,c}(t)\in[0, 1]$, and $\upsilon_{k,c}(t)=1$ means that IoT device $k$ requests task $c$ at time slot $t$, otherwise $\upsilon_{k,c}(t)=0$. Besides, $\varepsilon_m(t)\in[0, 1]$, and $\varepsilon_m(t)=1$ means that leader UAV $m$ returns to the depot at time slot $t$, otherwise $\varepsilon_m(t)=0$. Meanwhile, $\delta_{m,n}(t)\in[0, 1]$, and $\delta_{m,n}(t)=1$ means that follower UAV $n$ follows the leader UAV $m$ at time slot $t$, otherwise $\delta_{m,n}(t)=0$. Additionally, $\phi_{m,n,c}(t)\in[0, 1]$, and $\phi_{m,n,c}(t)=1$ means that follower UAV $n$ delegates task $c$ to the leader UAV $m$, otherwise $\phi_{m,n,c}(t)=0$.

We assume that $T^{off}_{n,k,c}(t)<T_n^{Fhov}(t)$, indicating that the hovering time $T_n^{Fhov}(t)$ is sufficiently long for follower UAV $n$ to receive each task offloaded by IoT devices at time slot $t$. Besides, the application $c$ placed in follower UAV $n$ and leader UAV $m$ can be defined as $\omega_{n,c}^F(t)\in\{0,1\}$ and $\omega_{m,c}^L(t)\in\{0,1\}$, respectively. $\omega_{n,c}^F(t)=1$ means that follower UAV $n$ place the application which can process task $c$, otherwise $\omega_{n,c}^F(t)=0$. And the definition of $\omega_{m,c}^L(t)$ is similar to that of $\omega_{n,c}^F(t)$. Consequently, the size of tasks processed by follower UAV $n$ and leader UAV $m$ can be expressed as:
\begin{equation}
	\begin{array}{l}
		Task^F_{n}(t)=\\
		\min\{\sum\limits_{k\in\mathcal{G}_{n}}\sum\limits_{m=1}^M\sum\limits_{c=1}^C(1-\phi_{m,n,c}(t))\upsilon_{k,c}(t)\omega^F_{n,c}(t)\kappa_{k,c},\\
		f_n^F(T_n^{Fhov}(t)-\min\{\bm{T}^{off}_n(t)\})\},
	\end{array}
\end{equation}
and
\begin{equation}
	\begin{array}{l}
		Task^L_{m}(t)=\\
		\min\{\sum\limits_{n=1}^N\sum\limits_{c=1}^C\delta_{m,n}(t)\phi_{m,n,c}(t)\upsilon_{k,c}(t)\omega^L_{m,c}(t)\kappa_{k,c},\\
		f_m^L(T_n^{Fhov}(t)-\min\{\bm{T}^{dele}_m(t)\})\},
	\end{array}
\end{equation}
respectively, where \resizebox{0.65\hsize}{!}{$\bm{T}^{off}_n(t)=\{T^{off}_{n,1,1}(t),...,T^{off}_{n,k,c}(t),...\}$} and \resizebox{0.65\hsize}{!}{$\bm{T}^{dele}_m(t)=\{T^{dele}_{m,1,1}(t),..., T^{dele}_{m,n,c}(t),...\}$}. $f_n^F$ and $f_m^L$ indicate the computing capacity of follower UAV $n$ and leader UAV $m$ (the number of CPU cycles per second). Then, the energy consumption of follower UAV $n$ and leader UAV $m$ computing tasks can be respectively written as $E_n^{comp}(t)=\xi(f^F_n)^2Task^F_n(t)$ and $E_m^{comp}(t)=(1-\varepsilon_m(t))\xi(f^L_m)^2Task^L_m(t)$, where $\xi$ denotes effective capacitance coefficient. Moreover, the propulsion energy consumption of leader UAV $m$ and follower UAV $n$ with the velocity $v$ can be expressed as:
\begin{equation}
	E_m^{pro}(t)=(1-\varepsilon_m(t))(P^{pro}(v)\frac{l}{v}+P^{pro}(0)T_n^{Lhov}(t)), 
\end{equation}
and
\begin{equation}
\begin{array}{l}
\resizebox{0.98\hsize}{!}{$E_n^{pro}(t)=P^{pro}(v)(\frac{((1-\delta_{m,n}(t-1)\delta_{m,n}(t))d_{m,n}(t)}{v}+\frac{q}{v})$}\\
+P^{pro}(0)T_n^{Fhov}(t), 
\end{array}
\end{equation}
respectively, where $P^{pro}(v)$ is the propulsion power model of UAVs, and its description follows \cite{36}. Furthermore, let $E^{ret}_m (t)$ be its energy consumption of returning to the depot, which can be written as $E^{ret}_m(t)=\varepsilon_m(t)P^{pro}(v)d_m^{ret}(t)/v$, where $d_m^{ret}(t)$ indicates the distance between leader UAV $m$ and the depot at time slot $t$.


\subsection{Problem Formulation}
In this paper, the energy efficiency $E^{effi}(t)$ means the amount of tasks processed by all UAVs relative to their energy consumption during time slot $t$. This can be mathematically expressed as
\begin{equation}
\begin{array}{l}
	E^{effi}(t)=\\
	\resizebox{0.97\hsize}{!}{$\frac{\sum\limits_{m=1}^M Task^L_m(t)+\sum\limits_{n=1}^N Task^F_n(t)}{\sum\limits_{m=1}^M((E_m^{comp}(t)+E_m^{pro}(t)+E^{ret}_m(t))+\sum\limits_{n=1}^N(E_n^{comp}(t)+E_n^{pro}(t))}$}.\nonumber
\end{array}
\end{equation}

Then, in this paper, we aim to jointly optimize the dynamic clustering and scheduling of considered the UAV swarm assisted MEC system, with the objective of maximizing the long-term energy efficiency, and which can be formulated as 
\begin{align}
	&[\mathcal{P}1]:\max\limits_{\bm{\mathcal{L}}(t),\bm{\mathcal{F}}(t),\bm{\omega}^L(t),\bm{\varepsilon}(t),\bm{\delta}(t),\bm{\phi}(t)}\lim_{T\rightarrow+\infty}\frac{1}{T}\sum\limits_{t=1}^TE^{effi}(t)\nonumber\\
	s.t.,\;&\sum\nolimits_{m=1}^M\delta_{m,n}=1,\forall n\in\mathcal{N},\label{delta1}\\
	&\sum\nolimits_{n=1}^N\delta_{m,n}\geq 1,\forall m\in\mathcal{M},\label{delta2}\\
	&\sum\limits\nolimits_{m=1}^M\omega_{m,c}^L(t)\varepsilon_m(t)\geq 1, \forall c\in\mathcal{C},\label{QoS}\\
	&\lvert\mathcal{F}_n(t)-\mathcal{F}_n(t-1)\rvert^2=q^2, \forall n\in\mathcal{N}\label{Ft},\\
	&(1-\varepsilon_m(t))\lvert\mathcal{L}_m(t)-\mathcal{L}_m(t-1)\rvert^2=l^2, \label{Lt}\\
	&\lvert\mathcal{F}_n(t)-\mathcal{F}_{n'}(t)\rvert^2\geq q^2, \forall n'\in\mathcal{N}\backslash n\label{FF'},\\
	&\lvert\mathcal{L}_m(t)-\mathcal{L}_{m'}(t)\rvert^2\geq l^2, \forall m'\in\mathcal{M}\backslash m\label{LL'},\\
	&\delta_{m,n}(t)(1-\varepsilon_m(t))\lvert x_m^L(t)-x_n^F(t)\rvert\leq\frac{l-q}{2}, \label{LFx}\\
	&\delta_{m,n}(t)(1-\varepsilon_m(t))\lvert y_m^L(t)-y_n^F(t)\rvert\leq\frac{l-q}{2}, \label{LFy}
\end{align}
where $\bm{\mathcal{L}}(t)=\{\mathcal{L}_1(t),\mathcal{L}_2(t),...,\mathcal{L}_M(t)\}$ and $\bm{\mathcal{F}}(t)=$ $\{\mathcal{F}_1(t),$\\$\mathcal{F}_2(t),...,\mathcal{F}_N(t)\}$ denote the location sets of leader UAVs and follower UAVs at time slot $t$, respectively. Therein, $\mathcal{L}_m(t)=(x_m^L(t),y_m^L(t))$ and $\mathcal{F}_n(t)=(x_n^F(t),y_n^F(t))$ represent their horizontal coordinates at time slot $t$, respectively. Constraint (\ref{delta1}) indicates that each follower UAV must follow a leader UAV at each time slot; constraint (\ref{delta2}) indicates that each UAV swarm contains at least one follower UAV at each time slot; constraint (\ref{QoS}) indicates that each type of application should be installed in at least one leader UAV hovering over the target region at each time slot $t$; constraint (\ref{Ft}) and (\ref{Lt}) imply that each leader UAV and follower UAV can only move to the centers of large grid and small grid, respectively; constraint (\ref{FF'}) and (\ref{LL'}) indicate that each small grid and large grid can only be covered by one follower UAV and one leader UAV, respectively, in order to avoid potential collisions; constraint (\ref{LFx}) and (\ref{LFy}) imply that each follower UAV must remain within its UAV swarm. In the following sections, we will first analyze problem $[\mathcal{P}1]$, and then propose a novel algorithm to derive the corresponding solution.

\section{Problem Reformulation and Solution}\label{SectionPR}
\subsection{Problem Reformulation}
Considering the intelligence of UAVs, to solve problem $[\mathcal{P}1]$, we allow each UAV to autonomously make decisions while ensuring appropriate regulation of cooperation and competition among them. The objective is maximizing the energy efficiency of the UAV swarm assisted MEC system through dynamic clustering and scheduling their energy replenishment, application placement, trajectory planning and task delegation. Additionally, considering the uncertainty of the future environment information, for example, task requirements of IoT devices are not available in advance for UAVs, we reformulate the joint optimization problem $[\mathcal{P}1]$ as a series of strongly coupled multi-agent stochastic games as follows.

The coupled multi-agent stochastic games are ERSG $\langle \mathcal{U},$$\mathcal{S}^{ER},$$\mathcal{A}^{ER},$$\mathcal{P}^{ER},$$\mathcal{R}^{ER}\rangle$, APSG $\langle \mathcal{U},$$\mathcal{S}^{AP},$$\mathcal{A}^{AP},$$\mathcal{P}^{AP},$$\mathcal{R}^{AP}\rangle,$ \\LTSG $\langle \mathcal{U},$$\mathcal{S}^{LT},$$\mathcal{A}^{LT},$$\mathcal{P}^{LT},$$\mathcal{R}^{LT}\rangle,$ DCSG $\langle \mathcal{U},$$\mathcal{S}^{DC},$$\mathcal{A}^{DC},$$\mathcal{P}^{DC},$\\$\mathcal{R}^{DC}\rangle,$ FTSG $\langle \mathcal{U},$$\mathcal{S}^{FT},\mathcal{A}^{FT},$$\mathcal{P}^{FT},$$\mathcal{R}^{FT}\rangle$ and TDSG $\langle \mathcal{U},$$\mathcal{S}^{TD},$\\$\mathcal{A}^{TD},$$\mathcal{P}^{TD},$$\mathcal{R}^{TD}\rangle$. Specifically, for ERSG, each leader UAV $u\in\mathcal{U}$ will choose an action individually based on the current environment state $s^{ER}(t)\in\mathcal{S}^{ER}$ at the beginning of each time slot $t$, and then form a joint action $\bm{a}^{ER}(t)\in\mathcal{A}^{ER}$. After executing the joint action, rewards will be obtained according to $\mathcal{R}^{ER}$, and the environment states will turn to next ones with $\mathcal{P}^{ER}$. The descriptions of APSG, LTSG, DCSG, FTSG and TDSG are similar to that of ERSG, and are omitted here for conciseness. In the following subsection, we propose a novel algorithm, called RLDC, to obtain equilibriums of these coupled multi-agent stochastic games.

\subsection{RLDC Algorithm}\label{SectionTL}
Since the transitions of states and actions in ERSG, APSG, LTSG, DCSG, FTSG and TDSG satisfy the Markov property, we characterize the strategic decision processes of each leader UAV and follower UAV by a series of respective Markov decision processes (MDPs) \cite{81}.

\textbf{MDP for each leader UAV in ERSG}: 

\emph{1) Environment state for each leader UAV in ERSG}: To reduce the size of the state space in ERSG, we divide the energy of leader UAVs into several levels. Specifically, the energy level of leader UAV $m$ can be written as $E^{level}_{m}(t)=\lceil E^{remain}_m/E^{unit}\rceil$, where $E^{unit}$ is the UAV energy unit. Hence, the environment state $s^{ER}(t) \in \mathcal{S}^{ER}$ for each leader UAV $m \in \mathcal{M}$ in ERSG at time slot $t$ can be expressed as $s^{ER}(t)=\bm{E}^{level}(t)$, where $\bm{E}^{level}(t)=\{E^{level}_1(t), E^{level}_2(t),..., E^{level}_M(t)\}$ indicates the set of all leader UAVs' energy levels. 

\emph{2) Action for each leader UAV in ERSG}: At time slot $t$, leader UAV $m \in \mathcal{M}$ chooses an action $a^{ER}_m(t) \in \mathcal{A}^{ER}_m$, where $\mathcal{A}^{ER}_m$ is the action set of UAV $m$ in ERSG consisting of two actions, i.e., return to the depot or not.

\emph{3) Reward of each leader UAV in ERSG}:
The immediate reward $r^{ER}_m(t)\in\mathcal{R}^{ER}_m$ of leader UAV $m \in \mathcal{M}$ at time slot $t$ is given by:
\begin{equation}
	r^{ER}_m(t)=\left\{\begin{array}{l}
		-10,\;\;\text{if constraint\;(\ref{QoS}) is violated},\\
		\varepsilon_m(t),\;\;\text{otherwise}.
	\end{array} \right. \label{r^{ER}}
\end{equation}

\emph{4) State transition probabilities of leader UAVs in ERSG}: The state transition probability from state $s^{ER}$ to state $s^{ER'}$ by taking the joint action $\bm{a}^{ER}(t) = (a_1^{ER}(t), a_2^{ER}(t), ..., a_M^{ER}(t))$ can be expressed as $\mathcal{P}^{ER}_{s^{ER}, s^{ER'}}(\bm{a}^{ER}(t)) = Pr(s^{ER}(t+1) = s^{ER'} | s^{ER}(t) = s^{ER}, \bm{a}^{ER}(t))$.

Note that, the transition probabilities of the other MDPs are similar to that in ERSG, and they are omitted subsequently for conciseness.

\textbf{MDP for each leader UAV in APSG}: 

\emph{1) Environment state for each leader UAV in APSG}: The environment state $s^{AP}(t) \in \mathcal{S}^{AP}$ for each leader UAV $m \in \mathcal{M}$ in APSG at time slot $t$ consists of applications placed in all leader UAVs, which can be expressed as $s^{AP}(t)=\bm{\omega}^L(t)$.

\emph{2) Action for each leader UAV in APSG}:
At time slot $t$, leader UAV $m \in \mathcal{M}$ chooses an action $a^{AP}_m(t) \in \mathcal{A}^{AP}_m$. $\mathcal{A}^{AP}_m$ signifies that the action set of leader UAV $m$ consisting of $C!/((C-S^L)*S^L!)$ actions, where $S^L$ denotes the maximum number of applications placed on leader UAV.

\emph{3) Reward of each leader UAV in APSG}:
The immediate reward $r^{AP}_m(t)\in\mathcal{R}^{AP}_m$ of leader UAV $m \in \mathcal{M}$ at time slot $t$ is given by:
\begin{equation}
	r^{AP}_m(t)=\sum_{\tau=1}^{t}\sum\limits_{n=1}^{N}\sum\limits_{k\in\mathcal{G}_n}\delta_{m,n}(\tau)\upsilon_k(\tau)\omega_m(\tau), \label{r^{AP}}
\end{equation}
where $r^{AP}_m(t)$ indicates the amount of tasks computed by leader UAV $m$ before time slot $t$.

\textbf{MDP for each leader UAV in LTSG}: 

\emph{1) Environment state for each leader UAV in LTSG}: The environment state $s^{LT}(t) \in \mathcal{S}^{LT}$ for each leader UAV $m \in \mathcal{M}$ in LTSG at time slot $t$ consists of all leader UAVs' positions $\bm{\mathcal{L}}(t)$ and set $\bm{\delta}(t)$, which can be expressed as $s^{LT}(t)=\{\bm{\mathcal{L}}(t), \bm{\delta}(t)\}$.

\emph{2) Action for each leader UAV in LTSG}:
At time slot $t$, leader UAV $m \in \mathcal{M}$ chooses an action $a^{LT}_m(t) \in \mathcal{A}^{LT}_m$, where $\mathcal{A}^{LT}_m$ is the action set of leader UAV $m$ in LTSG consisting of four possible actions, i.e., moving forward, backward, left or right to an adjacent large grid.

\emph{3) Reward of each leader UAV in LTSG}:
The immediate reward $r^{LT}_m(t)\in\mathcal{R}^{LT}_m$ of leader UAV $m \in \mathcal{M}$ at time slot $t$ is given by:
\begin{equation}
	r^{LT}_m(t)=\frac{Task^L_m(t)+\sum\limits_{n=1}^N \delta_{m,n}(t)Task^F_n(t)}{E^{remain}_m(t-1)-E^{remain}_m(t)}, \label{r^{LT}}
\end{equation}
where the numerator indicates the size of tasks computed by the UAV swarm, and the denominator represents the energy consumption of leader UAV $m$. 

\textbf{MDP for each follower UAV in DCSG}: 

\emph{1) Environment state for each follower UAV in DCSG}: The environment state $s^{DC}(t) \in \mathcal{S}^{DC}$ for each follower UAV $n \in \mathcal{N}$ at time slot $t$ consists of all leader UAVs' positions $\bm{\mathcal{L}}(t)$ and set $\bm{\delta}(t)$, which can be expressed as $s^{DC}(t)=\{\bm{\mathcal{L}}(t), \bm{\delta}(t)\}$.

\emph{2) Action for each follower UAV in DCSG}:
At time slot $t$, follower UAV $n \in \mathcal{N}$ chooses an action $a^{DC}_n(t) \in \mathcal{A}^{DC}_n$, where $\mathcal{A}^{DC}_n$ signifies the action set of follower UAV $n$ in DCSG consisting of $M$ possible actions.

\emph{3) Reward of each follower UAV in DCSG}:
The immediate reward $r^{DC}_n(t)\in\mathcal{R}^{DC}_n$ of follower UAV $n \in \mathcal{N}$ in DCSG at time slot $t$ is given by $r^{DC}_n(t)=E^{effi}(t)$.

\textbf{MDP for each follower UAV in FTSG}: 

\emph{1) Environment state for each follower UAV in FTSG}: The environment state $s^{FT}(t) \in \mathcal{S}^{FT}$ for each follower UAV $n \in \mathcal{N}$ in FTSG at time slot $t$ consists of all follower UAVs' positions $\bm{\mathcal{F}}(t)$ and set $\bm{\delta}(t)$, which can be expressed as $s^{FT}(t)=\{\bm{\mathcal{F}}(t), \bm{\delta}(t)\}$.

\emph{2) Action for each follower UAV in FTSG}:
At time slot $t$, follower UAV $n \in \mathcal{N}$ chooses an action $a^{FT}_n(t) \in \mathcal{A}^{FT}_n$, where $\mathcal{A}^{FT}_n(t)$ is the action set of follower UAV $n$ in FTSG consisting of four possible actions, i.e., moving forward, backward, left or right to an adjacent large grid.

\emph{3) Reward of each follower UAV in FTSG}:
The immediate reward $r^{FT}_n(t)\in\mathcal{R}^{FT}_n$ of follower UAV $n \in \mathcal{N}$ in FTSG at time slot $t$ is given by:
\begin{equation}
\begin{array}{l}
	\resizebox{0.98\hsize}{!}{$r^{FT}_n(t)=\frac{Task^F_n(t)}{\sum\limits_{m=1}^M\sum\limits_{c=1}^C\delta_{m,n}p_n^FT^{dele}_{m,n,c}(t)+\xi(f^F_n)^2Task^F_{n}(t)+E_n^{pro}(t)}, \label{r^{FT}}$}
\end{array}
\end{equation}
where the numerator indicates the size of tasks computed by the follower UAV $n$, and the denominator represents the energy consumption of follower UAV $n$. 

\textbf{MDP for each follower UAV in TDSG}: 

\emph{1) Environment state for each follower UAV in TDSG}: The environment state $s^{TD}(t) \in \mathcal{S}^{TD}$ for each follower UAV $n \in \mathcal{N}$ in TDSG at time slot $t$ consists of leader UAV $m$'s application placement $\bm{\omega}^L_m(t)$, follower UAV $n$'s application placement $\bm{\omega}^F_n(t)$ and set $\bm{\delta}(t)$, which can be expressed as $s^{TD}(t)=\{\bm{\omega}^L_m(t), \bm{\omega}^F_n(t), \bm{\delta}(t)\}$.

\emph{2) Action for each follower UAV in TDSG}:
At time slot $t$, follower UAV $n \in \mathcal{N}$ chooses an action $a^{TD}_n(t) \in \mathcal{A}^{TD}_n$, where $\mathcal{A}^{TD}_n(t)$ is the action set of follower UAV $n$ in TDSG consisting of two possible actions, i.e., whether delegating its tasks to the leader UAV or not.

\emph{3) Reward of each follower UAV in TDSG}:
The immediate reward $r^{TD}_n(t)\in\mathcal{R}^{TD}_n$ of follower UAV $n \in \mathcal{N}$ in TDSG at time slot $t$ is given by:
\begin{equation}
\begin{array}{l}
	r^{TD}_n(t)=\\
	\resizebox{0.98\hsize}{!}{$\sum\limits_{m=1}^M\sum\limits_{c=1}^C(\frac{Task^F_{n}(t)+\delta_{m,n}(t)Task^L_{m}(t)}{p_n^FT^{dele}_{m,n,c}(t)+\xi(f^F_n)^2Task^F_{n}(t)+\delta_{m,n}(t)\xi(f^L_m)^2Task^L_m(t)}), \label{r^{TD}}$}
\end{array}
\end{equation}
where the numerator indicates the size of tasks computed by leader UAV $m$ and follower UAV $n$ in the same swarm, and the denominator represents the energy consumption of task delegation and task computing.

Based on these MDPs, we propose a novel RLDC algorithm, where Q-learning is utilized to obtain the solution. RLDC includes a series of corresponding learners, namely, leader UAV energy replenishment learner, leader UAV application placement learner, leader UAV trajectory planning learner, follower UAV trajectory planning learner, follower UAV dynamic clustering learner and follower UAV task delegation learner.

The policy of leader UAV energy replenishment learner in UAV $m$ is expressed as $\pi_m^{ER}:\mathcal{S}^{ER}\longrightarrow \mathcal{A}_m^{ER}$, which signifies a probability distribution of actions $a_m^{ER} \in \mathcal{A}_m^{ER}$ in a given state $s^{ER}$. 

The Q function of the leader UAV energy replenishment learner in UAV $m$ is the expected reward by executing action $a_m^{ER} \in \mathcal{A}_m^{ER}$ in state $s^{ER} \in \mathcal{S}^{ER}$ under the given policy $\pi_m^{ER}$, which can be expressed by:
\begin{equation}
	\begin{array}{l}
		Q_m^{ER}(s^{ER},\bm{a}^{ER},\pi^{ER}_m)= \\
		\mathbb{E}(\sum\limits_{\tau=0}^{\infty}\sigma^{\tau}\mathcal{R}_m^{ER}(t+\tau+1)|s^{ER}(t) = s^{ER},\\
		\bm{a}(t)^{ER} = \bm{a}^{ER}, \pi_m^{ER}),\\
	\end{array}\label{QER}
\end{equation}
where $\sigma$ is a constant discounted factor with $\sigma \in[0, 1]$, and the value of (\ref{QER}) is termed as action value, i.e., Q value.

For striking a balance between exploration and exploitation, in this paper, we consider an $\epsilon$-greedy exploration strategy for the leader UAV energy replenishment learner. Specifically, the leader UAV energy replenishment learner in UAV $m$ selects a random action $a_m^{ER} \in \mathcal{A}_m^{ER}$ in state $s^{ER} \in \mathcal{S}^{ER}$ with probability $\epsilon$, and selects the best action $a_m^{ER*}$ with probability $(1-\epsilon)$, where the best action has $Q_m^{ER}(s^{ER},\bm{a}^{ER*},\pi^{ER}_m) \geq Q_m^{ER}(s^{ER},\bm{a}^{ER},\pi^{ER}_m)$, $\forall \bm{a}^{ER} \in \mathcal{A}^{ER}$ with $a_m^{ER*}$ being the $m$-th element of $\bm{a}^{ER*}$. Then, the probability of selecting action $a_m^{ER} \in \mathcal{A}_m^{ER}$ in state $s^{ER}$ can be expressed by:
\begin{equation}
	\begin{array}{l}
		\pi^{ER}_m(s^{ER},a^{ER}_m)\\
		=\left\{\begin{array}{l}
			1-\epsilon, \text{if $Q^{ER}_m(s^{ER},\cdot,\cdot)$ of $a^{ER}_m$ is the highest},\\
			\epsilon, \text{otherwise.}
		\end{array} \right.
	\end{array}
\end{equation}

In the Q value update step of Q-learning, the leader UAV energy replenishment learner follows the update rule:
\begin{equation}
	\begin{array}{l}
		Q_m^{ER}(s^{ER},\bm{a}^{ER},t+1)= \\
		Q_m^{ER}(s^{ER},\bm{a}^{ER},t)
		+\beta^{ER}(\mathcal{R}_m^{ER}(t)+ \\
		\max\limits_{\bm{a}^{{ER}'}\in \mathcal{A}^{ER}}\sigma Q_m^{ER}(s^{{ER}'},\bm{a}^{{ER}'},t)-Q_m^{ER}(s^{ER},\bm{a}^{ER},t)),
	\end{array}\label{QERupdate}
\end{equation}
where $\beta^{ER}$ denotes the learning rate of the leader UAV energy replenishment learner.

Since the settings of other learners are similar to those of leader UAV energy replenishment learner, the settings of other learners are omitted here for conciseness.

Overall, the RLDC algorithm is illustrated in Algorithm 1.
\begin{algorithm}[t]
	\footnotesize
	\caption{RLDC Algorithm} \label{algorithm1}
	Initialize Q value: $Q^{ER}_m=Q^{AP}_m=Q^{LT}_m=Q^{DC}_n=Q^{ER}_n=Q^{TD}_n=0$,
	$\forall m\in\mathcal{M},n\in\mathcal{N}$.\\
	Set the maximal iteration counter $LOOP$, $loop=0$ and $sum=0$.\\
	\For{$loop < LOOP$}{
		Set $t = 0$.\\
		\While{$t\leq T$}{
		    \For{$m=1$ to $M$}{
		        Observe state $s^{ER}(t)$, $s^{AP}(t)$, $s^{LT}(t)$.\\
			   Select $a^{ER}_m(t)$ according to $\pi^{ER}_m(s^{ER},\cdot)$.\\
			   \If {$\varepsilon_m(t)=1$}{
				   Select $a^{AP}_m(t)$ according to $\pi^{AP}_m(s^{AP},\cdot)$.}
			   \Else {Select $a^{LT}_m(t)$ according to $\pi^{LT}_m(s^{LT},\cdot)$.}
			}
			\For{$n=1$ to $N$}{
		         Observe state $s^{DC}(t)$, $s^{FT}(t)$, $s^{TD}(t)$.\\
			    Select $a^{DC}_n(t)$ according to $\pi^{DC}_n(s^{DC},\cdot)$.\\
			    Select $a^{FT}_n(t)$ according to $\pi^{FT}_n(s^{FT},\cdot)$.\\
			    Select $a^{TD}_n(t)$ according to $\pi^{TD}_n(s^{TD},\cdot)$.\\
			}
			Obtain the $E^{effi}(t)$ and the rewards $\mathcal{R}^{ER}_m(t)$, $\mathcal{R}^{AP}_m(t)$, $\mathcal{R}^{LT}_m(t)$, $\mathcal{R}^{DC}_n(t)$, $\mathcal{R}^{FT}_n(t)$ and $\mathcal{R}^{TD}_n(t)$.\\
			Update the Q values $Q^{ER}_m(t)$, $Q^{AP}_m(t)$, $Q^{LT}_m(t)$, $Q^{DC}_n(t)$, $Q^{FT}_n(t)$ and $Q^{TD}_n(t)$.\\
			Set $t=t+1$.
		}
		Set $sum=sum+\sum_{t=1}^TE^{effi}(t)$\\
		Set $loop=loop+1$.
	}
      Output: $sum/loop$
\end{algorithm}

\begin{table}[!t]
	\caption{Simulation Parameters} \label{Parameters}
	\footnotesize
	\begin{tabular}{p{20pt}|p{20pt}|p{20pt}|p{36pt}|p{23pt}|p{56pt}}
		
		\hline\noalign{\smallskip}
		\textbf{Param.} & \textbf{Value} & \textbf{Param.} & \textbf{Value}& \textbf{Param.} & \textbf{Value}\\
		\noalign{\smallskip}\hline\noalign{\smallskip}
		$M$ & $3$ &  $H_L$ & $150\mathrm{m}$ & $p^L$ & $2\mathrm{W}$\\
		$N$& $9$ &  $H_F$ & $120\mathrm{m}$ & $p^F$ & $0.2\mathrm{W}$ \\
		$K$ & $500$ &  $t$ & $30\mathrm{s}$  & $f^L$ & $2\mathrm{Mbps}$   \\
		$C$ &$10$ & $v$ & $20\mathrm{m/s}$ & $f^F$ & $2\mathrm{Mbps}$   \\
		$S_L$  & $6$ & $B$ & $10\mathrm{MHz}$  &  $q$ & $100\mathrm{m}$   \\
		$S_F$ & $4$ & $f$ & $3\mathrm{GHz}$ & $\varpi$ & $-174\mathrm{dBm/Hz}$ \\
		$\xi$ & $10^{-18}$ & $\kappa_{k,c}$ &  $10\mathrm{Mbits}$ & Target region & $2500\mathrm{m}\times2500\mathrm{m}$\\
		\hline
	\end{tabular}
\end{table}
\begin{figure*}[!t]
	\centering
	\addtocounter{figure}{-1}
	\label{fig:ThreeFig}
	\subfigure{
		\begin{minipage}[t]{0.3\linewidth}
			\centering
			\includegraphics[height=1.9in, width=2.4in]{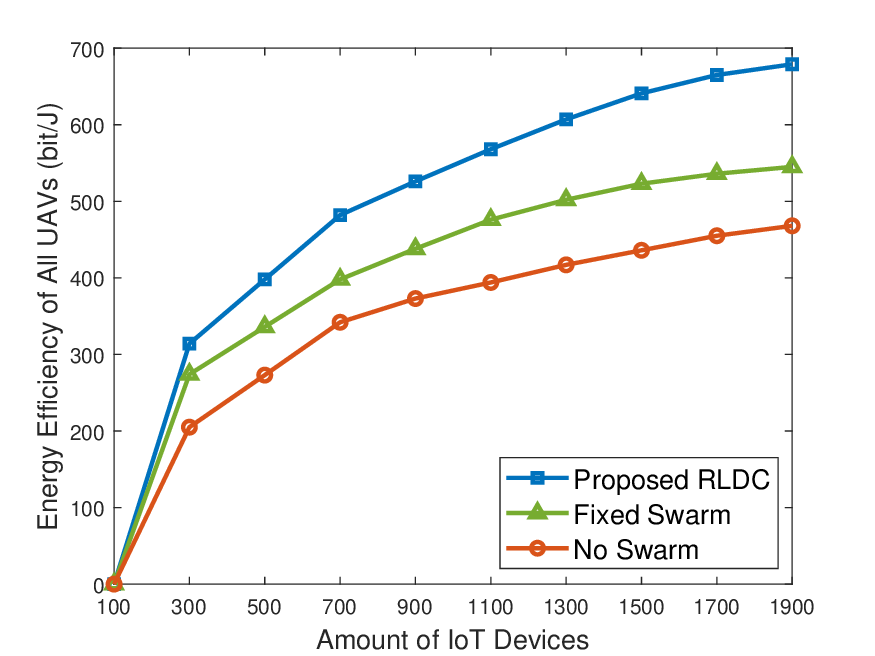}
			\caption{Energy efficiency w.r.t. the amounts of IoT devices.}\label{simu1}
		\end{minipage}%
	}%
	\hspace{1mm}
	\subfigure{
		\begin{minipage}[t]{0.3\linewidth}
			\centering
			\includegraphics[height=1.9in, width=2.4in]{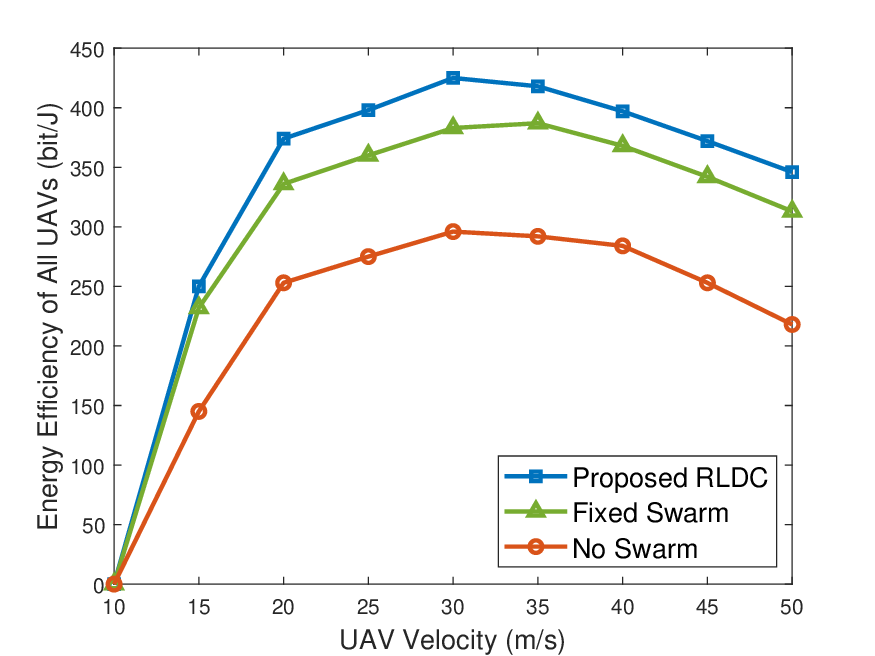}
			\caption{Energy efficiency w.r.t. UAV velocities.}\label{simu2}
		\end{minipage}%
	}%
	\hspace{1mm}
	\subfigure{
		\begin{minipage}[t]{0.3\linewidth}
			\centering
			\includegraphics[height=1.9in, width=2.4in]{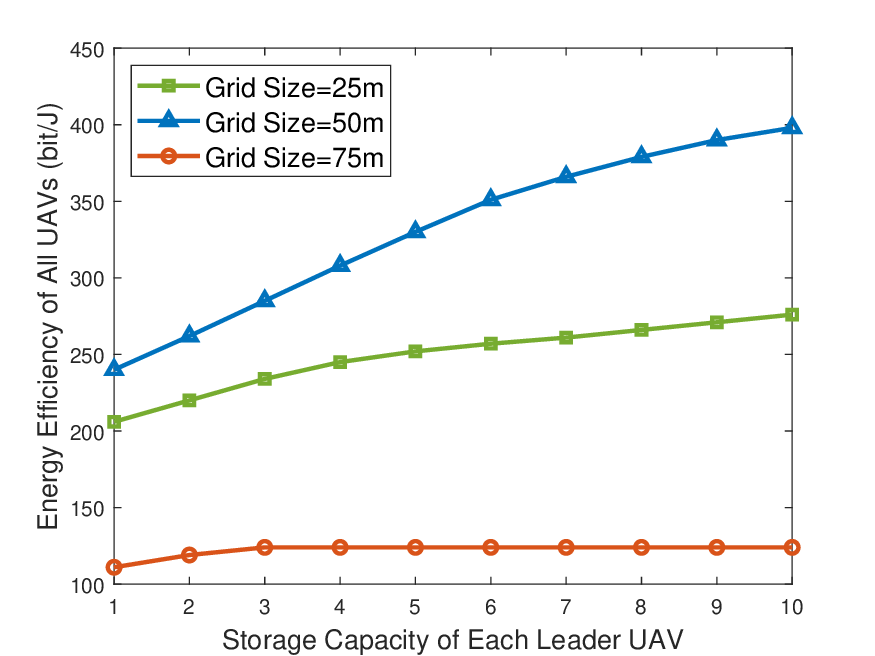}
			\caption{Energy efficiency w.r.t. storage capacities of each leader UAV.}\label{simu3}
		\end{minipage}
	}%
	\centering
\end{figure*}

\section{SIMULATION RESULTS}\label{SectionSIM}
In this section, we conduct extensive simulations to evaluate the performance of the proposed RLDC algorithm. The values of all simulation parameters are listed in Table \ref{Parameters}. Similar settings have also been utilized in previous work \cite{72,45}. 

For the purpose of comparison, we introduce two benchmark algorithms: a fixed UAV swarm algorithm and a no UAV swarm algorithm. Fixed swarm algorithm is originally designed to maximize the energy efficiency of all UAVs without considering dynamic clustering based on RLDC algorithm. No UAV swarm algorithm is originally designed to maximize the energy efficiency of all UAVs without considering UAV swarms based on RLDC algorithm.

Fig. 2 examines the energy efficiency of all UAVs as the amount of IoT devices varies. Obviously, the energy efficiency of all UAVs exhibits a monotonically increasing trend with the increasing number of IoT devices. This can be attributed to the generation of more task requests by IoT devices as their quantity grows. Furthermore, the results also demonstrate that the proposed RLDC algorithm surpasses both the fixed UAV swarm and no UAV swarm algorithm. This superiority arises from several reasons: i) in cases where the task requests from IoT devices are dynamically changing, the fixed UAV swarm can not dynamically cluster according to the ever-changing task requests, and thereby, many tasks can not be processed; ii) the storage capacities of each UAV are limited, and furthermore, they are unable to delegate the incapable tasks to other UAVs.

Fig. 3 illustrates the energy efficiency of all UAVs as the UAV velocity varies. Obviously, the energy efficiency of all UAVs initially increases and then decreases with the UAV velocity increasing. Because as the velocity of UAVs increases, there is a reduction in the time taken for movement. As a result, UAVs have more time available for hovering and processing tasks. However, the increase in the number of processing tasks is offset by the higher energy consumption of UAVs caused by their fast velocity. As a result, the drawbacks ultimately outweigh the benefits. The explanations of the proposed RLDC algorithm outperforms the other two algorithms are consistent with those discussed in Fig. 2.

Fig. 4 demonstrates the energy efficiency of all UAVs with varying storage capacities for each leader UAV. It can be observed that the performance under grid size 50$m$ outperforms 25$m$ and 75$m$. The reason is that grid size 25$m$ includes fewer IoT devices, leading to decreased number of tasks processed by UAVs. In contrast, while the grid size of 75$m$ may contain more IoT devices, it also results in a substantial increase in the energy consumption of UAVs during movement. Therefore, among these three sizes, the grid size of 50$m$ is the most suitable. Furthermore, the results also indicate that the energy efficiency of all UAVs increases as the storage capacity of each UAV grows, which can be attributed to the increased ability to process more types of applications for each UAV.

\section{Conclusion}\label{SectionCON}

In this paper, with the aim of maximizing the long-term energy efficiency of the UAV swarm assisted MEC system, a joint optimization problem of UAVs' dynamic clustering and scheduling is formulated. By taking into account the inherent cooperation and competition among intelligent UAVs, we reformulate the optimization problem as a series of coupled multi-agent stochastic games, and then propose a novel RLDC algorithm for obtaining equilibriums. Simulation results show that, compared to counterparts, the proposed RLDC algorithm can significantly increase the energy efficiency of the UAV swarm assisted MEC system.
\vspace{2ex}


\bibliographystyle{IEEEtran}
\bibliography{IEEEabrv,ref}

\begin{thebibliography}{10}
\providecommand{\url}[1]{#1}
\csname url@samestyle\endcsname
\providecommand{\newblock}{\relax}
\providecommand{\bibinfo}[2]{#2}
\providecommand{\BIBentrySTDinterwordspacing}{\spaceskip=0pt\relax}
\providecommand{\BIBentryALTinterwordstretchfactor}{4}
\providecommand{\BIBentryALTinterwordspacing}{\spaceskip=\fontdimen2\font plus
\BIBentryALTinterwordstretchfactor\fontdimen3\font minus
  \fontdimen4\font\relax}
\providecommand{\BIBforeignlanguage}[2]{{%
\expandafter\ifx\csname l@#1\endcsname\relax
\typeout{** WARNING: IEEEtran.bst: No hyphenation pattern has been}%
\typeout{** loaded for the language `#1'. Using the pattern for}%
\typeout{** the default language instead.}%
\else
\language=\csname l@#1\endcsname
\fi
#2}}
\providecommand{\BIBdecl}{\relax}
\BIBdecl

\bibitem{57}
Y.~Liao, X.~Chen, S.~Xia, Q.~Ai, and Q.~Liu, ``Energy minimization for {UAV}
  swarm-enabled wireless inland ship {MEC} network with time windows,''
  \emph{IEEE Trans. Green Commun. Netw.}, vol.~7, no.~2, pp. 594--608, June
  2023.

\bibitem{53}
Y.~Shi, C.~Yi, R.~Wang, Q.~Wu, B.~Chen, and J.~Cai, ``Service migration or task
  rerouting: A two-timescale online resource optimization for {MEC},''
  \emph{IEEE Trans. Wirel. Commun.}, pp. 1--1, July 2023.

\bibitem{67}
W.~He, H.~Yao, T.~Mai, F.~Wang, and M.~Guizani, ``Three-stage stackelberg game
  enabled clustered federated learning in heterogeneous {UAV} swarms,''
  \emph{IEEE Trans. Veh. Technol.}, vol.~72, no.~7, pp. 9366--9380, July 2023.

\bibitem{4}
C.~Zhan and Y.~Zeng, ``Completion time minimization for multi-{UAV}-enabled
  data collection,'' \emph{IEEE Trans. Wireless Commun.}, vol.~18, no.~10, pp.
  4859--4872, Oct. 2019.

\bibitem{60}
J.~Chen, C.~Yi, J.~Li, K.~Zhu, and J.~Cai, ``A triple learner based energy
  efficient scheduling for multi-{UAV} assisted mobile edge computing,'' in
  \emph{Proc. {IEEE ICC}}, Jun. 2023.

\bibitem{61}
K.~Wang, X.~Zhang, L.~Duan, and J.~Tie, ``Multi-{UAV} cooperative trajectory
  for servicing dynamic demands and charging battery,'' \emph{IEEE Trans. Mob.
  Comput.}, vol.~22, no.~3, pp. 1599--1614, Mar. 2023.

\bibitem{62}
T.~Li, S.~Leng, Z.~Wang, K.~Zhang, and L.~Zhou, ``Intelligent resource
  allocation schemes for {UAV}-swarm-based cooperative sensing,'' \emph{IEEE
  Internet Things J.}, vol.~9, no.~21, pp. 21\,570--21\,582, Nov. 2022.

\bibitem{75}
A.~Mukherjee, S.~Misra, V.~S.~P. Chandra \emph{et~al.}, ``Resource-optimized
  multiarmed bandit-based offload path selection in edge {UAV} swarms,''
  \emph{IEEE Internet Things J.}, vol.~6, no.~3, pp. 4889--4896, June 2019.

\bibitem{43}
J.~Li, C.~Yi, J.~Chen, K.~Zhu, and J.~Cai, ``Joint trajectory planning,
  application placement, and energy renewal for {UAV}-assisted {MEC}: A
  triple-learner-based approach,'' \emph{IEEE Internet of Things J.}, vol.~10,
  no.~15, pp. 13\,622--13\,636, Aug. 2023.

\bibitem{36}
Y.~Zeng, J.~Xu, and R.~Zhang, ``Energy minimization for wireless communication
  with rotary-wing {UAV},'' \emph{IEEE Trans. Wireless Commun.}, vol.~18,
  no.~4, p. 2329–2345, Apr. 2019.

\bibitem{81}
R.~Chen, C.~Yi, K.~Zhu, B.~Chen, J.~Cai, and M.~Guizani, ``A three-party
  hierarchical game for physical layer security aware wireless communications
  with dynamic trilateral coalitions,'' \emph{IEEE Trans. Wirel. Commun.}, Oct.
  2023.

\bibitem{72}
C.~Zhao, J.~Liu, M.~Sheng, W.~Teng, Y.~Zheng, and J.~Li, ``Multi-{UAV}
  trajectory planning for energy-efficient content coverage: A decentralized
  learning-based approach,'' \emph{IEEE J. Sel. Areas Commun.}, vol.~39,
  no.~10, pp. 3193--3207, Oct. 2021.

\bibitem{45}
B.~Liu, Y.~Wan, F.~Zhou, Q.~Wu, and R.~Hu, ``Resource allocation and trajectory
  design for {MISO} {UAV}-assisted {MEC} networks,'' \emph{IEEE Trans. Veh.
  Technol.}, vol.~71, no.~5, pp. 4933--4948, May 2022.

\end{thebibliography}

\end{document}